\documentclass[11pt]{article}
\usepackage[dvips]{graphicx}
\newcommand{\cc}{\cite}

\newcommand{\be}{\begin{equation}}
\newcommand{\ee}{\end{equation}}

\def\ve{\varepsilon}
\def\w{\omega}

\def\ps{\psi}
\def\bps{\bar\psi}

\def\pd{\partial}

\def\L{\Lambda}

\def\z{\zeta}

\def\<{\langle}
\def\>{\rangle}

\def\Log{\hbox{ln}}

\def\g{\gamma}  
\def\d{\delta}  
\def\l{\lambda}   \def\L{\Lambda}

\def\m{\mu}
\def\n{\nu}

\def\z{\zeta}

\def\v{\vec}

\def\vf{\varphi}
\def\({\left(}
\def\[{\left[}
\def\){\right)}
\def\]{\right]}

\def\pd{\partial}
\def\dkt{{d^3 k \over (2\pi)^3}}
\def\dkf{{d^4 k \over (2\pi)^4}}
\def\dk{{d^n k \over (2\pi)^n}}

\def\cad{{\cal D}}
\def\w1{W^{(1)}}
\def\v1{V^{(1)}}

\begin{document}

\begin{center}

{\large \bf{Cosmological Constant and Zeta-Function}} \\

\vspace{5mm}

{Igor O.Cherednikov$^{a,b,}$\footnote{Igor.Cherednikov@jinr.ru}}
\\
 \vspace{2mm}
 $^a$ {
 \it Bogolyubov Laboratory of Theoretical Physics, JINR, 141980 Dubna,
 Russia\\
\vspace{0mm}
$^b$ Institute for Theoretical Problems of Microphysics, MSU, 119899 Moscow, Russia}
\\
\end{center}

\begin{center}
{ (27 Nov 2002)} \\
\end{center}

\begin{abstract}

\noindent
The relativistic invariant  zeta-function approach to computation of
the vacuum energy contribution to cosmological constant is discussed. It is shown that this
value is determined by the fourth power of the quantized field mass, while the dependence from the
large mass scale is only logarithmic.
This value is compared to the result obtained in the dimensional regularization scheme
which also satisfies the relativistic
invariance condition, and found to be the same up to irrelevant finite terms.
The consequences of the renormalization group invariance are also briefly discussed.
\end{abstract}
\vfill
\vspace{0mm}

\noindent
The cosmological constant contains the contributions of a various origin, the explicit evaluation
and ``fine tuning'' of which remains to be an open problem \cc{rev}. Here the only one of them
will be addressed --  the contribution of the ground state energy  of quantum fields.
The Einstein equation may be written in the form:
\be
R_{\m\n} - {1\over2} g_{\m\n} R = - g_{\m\n} \L_0- 8 \pi G T_{\m\n} \ ,
\label{ei} \ee where $\L$ is the ``classical'' part, while the vacuum average of the
energy-momentum tensor  $\<T_{\m\n}\>$ arises due to the quantum fluctuations. The condition of
the relativistic invariance for this quantity may be formulated
as \cc{rev, zel}:
\be
\<T_{\m\n}\> = \ve g_{\m\n} \ , \label{met}
\ee what means that the energy density $\ve = \<T_{00}\>$ and the vacuum pressure
$p = \<T_{ii}\>$ are related as:
\be
\ve = -p \ . \label{main}
\ee
Using (\ref{met}), Eq. (\ref{ei}) can be written in the form
\be R_{\m\n} - {1\over2} g_{\m\n} R = - g_{\m\n} \L_{eff} \ ,
\ee where
\be
\L_{eff} = \L_0 + 8 \pi G \ve \ , \label{den}
\ee is the effective cosmological term.
The usual estimation of the energy density for a scalar field is based on the obvious field
theoretical
formula with the UV cutoff of the divergent integral at the Planck scale $M_P \approx 10^{19} \ GeV$:
\be \ve = {1 \over 2} \ \int\! \dkt \ \omega_k  \to \int_0^{M_P}
 \! {k^2 dk \over (2\pi)^2} \sqrt{\hbox{\bf k}^2 + m^2} \ , \label{cut}\ee producing the
famous huge value \be \ve_P \approx 10^{71} \ GeV^4\ \ , \ee {\it i.e.}, about 120 orders larger
than the observable one.

However, recently it was argued in Ref. \cc{akh} that the leading power-law terms in this naive
evaluation do not satisfy the relativistic invariance condition (\ref{main}). Instead, it was
shown that the calculation of the zero-point energy in dimensional regularization
satisfies the condition (\ref{main}) at all steps, and yields after removing the singular part within
the $\overline{\hbox{MS}}$ scheme:
\be
\ve_{dim}^{(0)} = -p = -{m^4 \over 64 \pi^2} \(\Log {\L^2 \over m^2} + {3 \over 2}\) \ ,
\label{akh1}
\ee where $m$ is the field mass, and $\L$ is the mass scale parameter.\footnote{The similar
expression was found by A. D. Sakharov in Ref \cc{sakh} within the different approach. Analogous
situation is considered in Ref \cc{shap}.}
The dimensionally regularized, but non-renormalized expression contains neither quartic, nor quadratic
divergent terms -- the only logarithmic ones appear.
In contrast, the four-dimensional cutoff performed for computation the vacuum averaged trace of the
energy-momentum tensor does give the quadratic divergent term, being in the same time
relativistic invariant. In all these cases, it was shown that the vacuum
energy density  vanishes for a massless field \cc{akh}.

On the other hand, a powerful method for calculation of the ground state energy for various
configurations of quantum fields is provided by the $\z$-function technique supplied, if necessary,
with the heat-kernel expansion \cc{haw, dow, bor}.
Within the general framework of the zero-point energy computations, the classical cosmological term
$\L_0$ is treated as a ``bare constant'', and possible divergences appearing in calculation of the
quantum contribution are to be absorbed by the (infinite) redefinition of this  parameter.
One of the aims of the present paper is
to check if this  approach can be applied effectively, in a relativistic invariant manner,
to the cosmological constant calculations, and compare its results with
that ones obtained in the other invariant framework.

It can be easily shown, that the straightforward application of the $\z$-function regularization:
\be
\int \!\dkt \ \omega_k \to \int \!\dkt \ \omega_k^{1-\epsilon} \
\ee
doesn't meet the relativistic invariance
condition (\ref{main}). In order to use this approach properly, let us consider the following
definition of the one-loop vacuum energy of the free quantum field \cc{haw, brev,
pes}:
\be
\ve = {1 \over V_4} \Log Z(\vf) \ , \label{def1}
\ee where $V_4$ is the (Euclidean)
infinite space-time volume, the partition function for the scalar field $\vf(x)$ reads
\be
Z(\vf) = \int \! \cad \vf \exp\(- {i \over 2} \int \! dx \ \sqrt{-g}\vf(x) A_S \vf(x)\) \ , \label{zf}
\ee
and the second-order operator $A_S = g_{\m\n}\pd_\m\pd^\n + m^2$ has the eigenvalues $\l_k$:
\be
A_S \vf_k = \l_k \vf_k \ .
\ee The generalized dimensionless $\z$-function corresponding to this operator is defined as the
infinite sum
\be
\z (s) = \m^{2s}\sum_k \l_k^{-s} \ , \label{defz}
\ee where $\m$ is an arbitrary mass scale providing the correct dimension. In case of the
continuous spectrum $\l_k$ (which we're actually dealing with), the sum in (\ref{defz}) is replaced by the
integral with the corresponding measure
\be
\sum_n \to V_4 \int \! \dkf \ .
\ee
The function (\ref{defz}) converges in (3+1)D space-time for Re $(z) > 2$, and can be analytically
continued to a meromorphic function having the poles only at $s = 1$ and $s=2$ \cc{haw}. At the origin
($s = 0$) it is regular.

Then, from the expressions
\be
Z(\vf) = \(\hbox{det} \ A_S\)^{-1/2} = \exp \(\sum_k {m \over \l_k^{1/2}}\) \
\ , \ \ \Log Z(\vf) = -{1 \over 2} \sum_k \Log {\l_k \over m^2} \ ,
\ee evaluating the derivative of the $\z$-function at $s=0$
\be
\z'(0) = \sum_k \Log {\m^2 \over \l_k} \ ,
\ee one gets the formula for the zero-point energy
\be
\ve = {1 \over V_4} \Log Z(\vf) = {1 \over 2V_4} \(\z'(0) - \z(0) \Log {\m^2 \over m^2}\) \  \label{esk0}
\ee for the scalar field $\vf$ with the  mass $m$ (for comparison, see, {\it e.g.}, \cc{brev}).
Now, using the continuous spectrum of eigenvalues $\l_k$ in the momentum space
\be
\l_k = - k^2 + m^2 \ ,\ee we evaluate (replacing the infinite sum by the
corresponding integral and performing Wick rotation ) explicitly the
corresponding $\z$-function\footnote{By virtue of
simplicity of the expression for $\l_k$, we can do it. In general case, the eigenvalues are
unknown, and the heat-kernel technique should be applied \cc{haw, dow, bor}. In certain
situations, the direct large-$k$ expansion may also be useful \cc{ch2}.}:
\be
\z(s) = V_4 \m^{2s} \int \! \dkf {1 \over \(k^2 + m^2\)^s} = V_4 {m^4 \over 16 \pi^2}
\({\m^2 \over m^2}\)^s
{1 \over s^2 -3s +2} \ . \label{zf2}
\ee
Then we have
\be
\z(0) = {V_4 m^4 \over 32 \pi^2} \ \ , \ \ \z'(0) = {3 \over 4} {V_4 m^4 \over 16 \pi^2} \ ,
\ee and the energy density is given by
\be
\ve_\z = - {m^4 \over 64 \pi^2}  \( \Log {\m^2 \over m^2} - {3 \over 2} \) \ .
\label{esk3}
\ee
This value coincides with that one obtained in Ref. \cc{akh}  provided
that the arbitrary mass $\m$ is re-scaled to give the same log-independent term.
This coincidence may be formally explained as follows (for a comparison, see also Ref. \cc{gur}).
Computing the vacuum average of the traced
energy-momentum tensor
\be
\<0| T_{\m\m}|0\> = m^2 \<0|\vf^2|0\> \ ,
\ee  with account of the condition (\ref{main}) we find \cc{akh}:
\be
\ve = {m^2 \over 4} \int\! \dkf \ {i \over k^2 - m^2 +i\ve} \ . \label{pr1}
\ee
The equivalence of the definition (\ref{cut}) and the expression (\ref{pr1})
may be demonstrated, {\it e.g.}, if we
write from (\ref{cut}) and (\ref{main})
$$
\<0| T_{\m\m}|0\> = 4 \ve = \int\! \dkt {1 \over 2 \omega_k} \(\omega^2_k - \hbox{\bf k}^2\)
=$$ \be =
m^2 \int\!\dkf \ 2
\pi \ \d_+\(k^2 - m^2\) = m^2 \int\! \dkf \ {i \over k^2 - m^2 +i\ve} \ .  \label{dr2}
\ee
This integral can be calculated in the dimensional regularization scheme what gives:
\be
\ve_{dim}^{(1)} = - {m^4 \over 64 \pi^2}  \( {4\pi \m^2 \over m^2}\)^\ve
\( {1\over \ve} +1 - \g_E\)  \ . \label{dr3}
\ee
On the other hand, evaluating the vacuum energy density according the definition (\ref{def1})
in the same regularization scheme one gets:
\be
\ve_{dim}^{(2)} = -{1 \over 2} \ \int\! \dk \ \Log{m^2 \over k^2 -m^2 +i\ve} =
- {m^4 \over 64 \pi^2}  \( {4\pi \m^2 \over m^2}\)^\ve\( {1\over\ve} +{3\over2} - \g_E \) \ ,
\ee what coincides exactly
with the result obtained in Ref. \cc{akh} (in this paper, Eq.(\ref{akh1})) and differs by irrelevant
log-independent term with (\ref{dr3}). Therefore, one can see that  the dimensional and $\z$-function
regularization schemes satisfy the relativistic invariance condition and  yield the similar result.
In these cases, the power-law divergences are absent (in contrast to the
four-dimensional cutoff method, as well as any other relativistic invariant scheme with an UV
cutoff: here the quadratic term survives \cc{akh}), and the finite log-independent
difference between them can be eliminated by means of the corresponding re-scaling of the arbitrary mass $\m$.

The corresponding result for the fermion field $\ps(x)$ can also be obtained from the
fermionic partition function
\be
Z_F (\bps,  \ps) = \int \! \cad \bps \cad \ps \ \exp\(i\int\! dx \sqrt{-g}\bps(x) A_F \ps(x) \) \ ,
\ee  where
\be
A_F = i \hat\pd - m_f \ .
\ee
By means of the similar considerations, one gets the fermion field contribution to the zero-point
energy:
\be
\ve_f = {1 \over V_4} \Log Z_F (\bps, \ps) = -{4 \over V_4} \(\z'_f(0) - \z_f(0)\Log{\m \over m_f}\) \
\ee with the following relations between the scalar and fermion $\z$-functions:
\be
\z_f(0, m_f) = \z(0, m) \ \ , \ \ \z'_f(0, m_f) = {1 \over 2} \z'(0,m) \ ,
\ee and the factor of two added to take into account the anti-particle
states.
One sees now that the fermionic and scalar zero-point energy densities satisfy:
\be
\ve_f = - 4 \ve \ ,
\ee as it should be.

The expression for the vacuum energy (\ref{esk3}) has been obtained within the relativistic
covariant formulation from the very beginning, and we observe that the properly applied
$\z$-function regularization doesn't break the invariance. Note also, that within this approach
the expressions for the energy density are finite at all steps of calculations, containing no
divergent terms. This is the  well-known feature of the $\z$-function method, and follows directly
from the analytical properties of the generalized
$\z$-function \cc{haw, bor}.

The question remains to be answered is how to deal with the logarithmic dependence on the arbitrary
mass scale $\m$. One can choose, of course, to treat it as a high energy (UV) boundary of order of the
Planck scale $M_P$, but this is not a satisfactory way since the scale $\m$
is completely arbitrary and the fixing it at a certain value has not a
physical basis. Another possible way is to demand the total effective constant $\L_{eff}$ in
Eq. (\ref{den}) to be independent on this (now treating as ``unphysical'') scale parameter. This
requirement gives the renormalization invariance equation
\be
\m {d \over d\m } \L_{eff} = 0 \ , \label{gr}
\ee what means that the renormalized classical cosmological constant $\L_0$ (see Eq.(1)) becomes a ``running
constant'' having the logarithmic dependence on the scale
$\m$ \cc{shap, bb}:
\be
\L_0 (\m) = \L_0(\m_0) + {G m^4 \over 4\pi} \Log{\m^2 \over \m_0^2} \ ,
\ee
where the value $\L_0(\m_0)$ gives the boundary condition for the
solution of the differential equation (\ref{gr}). One may think that the
running parameter $\L_0(\m)$ depend on the initial value $B_0(\m_0)$ as well
as on $\m_0$ itself, but this is not the case. Indeed, it must not depend on the starting point,
what is provided by the renormalization invariance condition (\ref{gr}). Then
it's convenient to express the running constant in terms of a single
variable (thus excluding an extra parameter):
\be
M = \m_0 \exp\[-{4\pi \over G m^4} \L_0(\m_0)\] \ , \label{lam1}
\ee and write
\be
\L_0(\m) = {G m^4 \over 4 \pi} \Log {\m \over M} \ . \label{lam2}
\ee
The mass parameter $M$ plays now a role of the fundamental energy scale
analogous to $\L_{QCD}$, while the renormalized classical cosmological
constant $\L_0$ becomes a running parameter, the value of which can be
extracted from data at certain energy scale $\m$.
The expression for the renormalized effective cosmological constant then
reads:
\be
\L_{eff} = {G m^4 \over 4 \pi}\[\Log {m \over M } + {3 \over 2} \] \ ,
\label{eff1} \ee
and is controlled by the forth power of the (small) mass $m$. The dependence
on the experimentally detectable parameter $M$, which now can be identified with some high energy
scale, such as the Planck mass, is only logarithmic and affects slightly the
result.

To summarize, it is confirmed that the proper relativistic invariant calculation in a regularization
scheme which does not use an UV cutoff (such as dimensional, or $\z$-function regularization) of the
quantum zero-point energy contribution to the cosmological constant in the flat space-time
yields no power-law divergences
(just logarithmic ones in the dimensional regularization, and no divergences at all in the
$\z$-function regularization), and yields the result which is determined by the fourth power
of the elementary quanta mass rather that the large mass scale.
The dependence on the latter appears to be only logarithmic and
its influence on the result is not so important. In this simple study we
neglected the possible curvature of the space-time what might change the
results significantly. Also we did not take into account the possible
presence of different sorts of quantum fields which can contribute to the
vacuum energy \cc{shap}.

As compared to the other possible regularization schemes,
the $\z$-function method based on the formulas (\ref{def1},\ref{zf}) seems to have the
following advantages:

\noindent
(i) By virtue of the analytical properties of the generalized $\z$-function,
the divergences do not appear at any step of the calculations and hence, from the formal point of
view, no (infinite) renormalization is required.

\noindent
(ii) This method  starts with the covariant expression (\ref{zf}) which can be easily generalized to the
case of the curved space-time. It is important since the effects of the curvature may be
nontrivial\cc{haw, dow}.

\vspace{0.5cm}

\noindent
The author is most grateful to E. Kh. Akhmedov for fruitful discussions, critics and
useful explanations, and to H. Stefancic for discussion and pointing out
some errors in results.
This work is partially supported by RFBR (nos. 02-02-16194, 01-02-16431, 00-15-96577), INTAS
(no. 00-00-366), and Heisenberg-Landau program HL-2002-13.
The author thanks also the hospitality and financial support of
the CERN TH-Division where a major part of this was done.

\vspace{1cm}

\eject


\begin{thebibliography}{99}

\bibitem{rev} S. Weinberg, {\it Rev. Mod. Phys.} 61 (1989) 1; S. Weinberg, astro-ph/0005265;
A. Vilenkin, hep-th/0106083; V. Sahni, A. Starobinsky,
{\it Int. J. Mod. Phys. D} 9 (2000) 373.

\bibitem{zel} Ya. B. Zeldovich, {\it Sov. Phys. Uspekhi} 11 (1968) 381.

\bibitem{akh} E. Kh. Akhmedov, hep-th/0204048.

\bibitem{sakh} A. D. Sakharov, {\it Theor. Math. Phys.} 23 (1976) 435, translated from {\it Teor.
Mat. Fiz.} 23 (1975) 178 (in Russian).

\bibitem{shap} I. L. Shapiro, J. Sola, {\it Phys. Lett. B} 475 (2000) 236;
{\it JHEP} 02 (2002) 006.

\bibitem{haw} S. W. Hawking, {\it Commun. Math. Phys.} 55 (1977) 133.

\bibitem{dow} J. S. Dowker, R. Critchley, {\it Phys. Rev. D} 13 (1976) 3224.

\bibitem{bor} M. Bordag, U. Mohideen, V. M. Mostepanenko, {\it Phys. Reports} 353
(2001) 1.

\bibitem{brev} I. Brevik, L. N. Granda, S. O. Odintsov, {\it Phys. Lett.
B} 367 (1996) 206.

\bibitem{pes} M. E. Peskin, D. V. Schroeder, {\it An Introduction to Quantum Field
Theory} (Addison-Wesley Publishing Co., 1995).

\bibitem{ch2} I. Cherednikov, {\it Phys. Lett. B} 498 (2001) 40.

\bibitem{gur} V. G. Gurzadyan, S.-S. Xue, astro-ph/0105245.

\bibitem{bb} A. Babic, B. Guberina, R. Horvat, H. Stefancic, {\it Phys. Rev. D} 65 (2002) 085002.








\end{thebibliography}
\end{document}